\documentclass[preprint,12pt]{elsarticle}

\usepackage{graphics}
\usepackage{amssymb}
\usepackage{amsthm}

\usepackage{amsmath}
\DeclareMathOperator{\tr}{tr}

\DeclareMathOperator{\re}{Re}
\DeclareMathOperator{\im}{Im}
\DeclareMathOperator{\sign}{sign}

\makeatletter
  
  \@addtoreset{equation}{section}
\makeatother

\usepackage[dvips]{pict2e}

\begin{document}

\begin{frontmatter}

%% Title, authors and addresses

%% use the tnoteref command within \title for footnotes;
%% use the tnotetext command for the associated footnote;
%% use the fnref command within \author or \address for footnotes;
%% use the fntext command for the associated footnote;
%% use the corref command within \author for corresponding author footnotes;
%% use the cortext command for the associated footnote;
%% use the ead command for the email address,
%% and the form \ead[url] for the home page:
%%
%% \title{Title\tnoteref{label1}}
%% \tnotetext[label1]{}
%% \author{Name\corref{cor1}\fnref{label2}}
%% \ead{email address}
%% \ead[url]{home page}
%% \fntext[label2]{}
%% \cortext[cor1]{}
%% \address{Address\fnref{label3}}
%% \fntext[label3]{}

\title{Numerical simulation of the $\mathcal{N}=(2,2)$ Landau--Ginzburg model}

%% use optional labels to link authors explicitly to addresses:
%% \author[label1,label2]{<author name>}
%% \address[label1]{<address>}
%% \address[label2]{<address>}

\author{Syo Kamata}
\ead{skamata@rikkyo.ac.jp}

\address{Graduate School of Science, Rikkyo University,
3-34-1 Nishi-Ikebukuro, Toshima-ku, Tokyo 171-8501, Japan}

\author{Hiroshi Suzuki}
\ead{hsuzuki@riken.jp}

\address{%
%Quantum Hadron Physics Laboratory,
Theoretical Research Division, RIKEN Nishina Center, Wako 2-1, Saitama 351-0198,
Japan}

\begin{abstract}
%% Text of abstract
The two-dimensional $\mathcal{N}=(2,2)$ Wess--Zumino (WZ) model with a cubic
superpotential is numerically studied with a momentum-cutoff regularization
that preserves supersymmetry. A numerical algorithm based on the Nicolai map is
employed and the resulting configurations have no autocorrelation. This system
is believed to flow to an $\mathcal{N}=(2,2)$ superconformal field theory
(SCFT) in the infrared (IR), the $A_2$~model. From a finite-size scaling
analysis of the susceptibility of the scalar field in the WZ model, we
determine $1-h-\Bar{h}=0.616(25)(13)$ for the conformal dimensions~$h$
and~$\Bar{h}$, while $1-h-\Bar{h}=0.666\dots$ for the $A_2$~model. We also
measure the central charge in the IR region from a correlation function between
conserved supercurrents and obtain $c=1.09(14)(31)$ ($c=1$ for the
$A_2$~model). These results are consistent with the conjectured emergence of
the $A_2$~model, and at the same time demonstrate that numerical studies can be
complementary to analytical investigations for this two-dimensional
supersymmetric field theory.
\end{abstract}
% Report number: RIKEN-QHP-7
\begin{keyword}
%% keywords here, in the form: keyword \sep keyword
Supersymmetry\sep Non-perturbative study\sep Landau--Ginzburg model\sep%
Nicolai map
%% MSC codes here, in the form: \MSC code \sep code
%% or \MSC[2008] code \sep code (2000 is the default)

\end{keyword}

\end{frontmatter}

%%
%% Start line numbering here if you want
%%
% \linenumbers

%% main text
\section{Introduction}
It is believed that the infrared (IR) limit of the two-dimensional
$\mathcal{N}=(2,2)$ Wess--Zumino model\footnote{This system is obtained by a
dimensional reduction of the four-dimensional Wess--Zumino
model~\cite{Wess:1974tw} from four dimensions to two dimensions.} (2D
$\mathcal{N}=(2,2)$ WZ model) with a quasi-\hspace{0pt}homogeneous
superpotential\footnote{A polynomial $W(\Phi)$ of variables $\Phi_I$ ($I=1$,
$2$, \dots, $N$) is called quasi-homogeneous when there exist some
weights~$\omega_I$ such that $W(\Phi_I\to\Lambda^{\omega_I}\Phi_I)
=\Lambda W(\Phi)$.} is a non-trivial $\mathcal{N}=(2,2)$ superconformal field
theory (SCFT)~\cite{Kastor:1988ef,Vafa:1988uu,Lerche:1989uy,Howe:1989qr,%
Cecotti:1989jc,Howe:1989az,Cecotti:1989gv,Cecotti:1990kz,Witten:1993jg}. See
Section~19.4 of~Ref.~\cite{Polchinski:1998rr} and Section~14.4
of~Ref.~\cite{Hori:2003ic} for reviews. This Landau--Ginzburg (LG)
description~\cite{Zamolodchikov:1986db} of $\mathcal{N}=(2,2)$ SCFT is a
remarkable non-perturbative phenomenon in field theory and physically, for
example, provides a basis for application of the gauged linear sigma
model~\cite{Witten:1993yc} to the Calabi-Yau compactification. Although the
emergence of SCFT has been tested in various ways, it is very difficult to
confirm this phenomenon directly in correlation functions, because the 2D WZ
model is strongly coupled in low energies. Application of conventional
numerical techniques (such as the lattice) would not be straightforward either,
because supersymmetry (SUSY) must be essential in the above non-perturbative
dynamics.

In a recent interesting paper~\cite{Kawai:2010yj}, Kawai and Kikukawa revisited
this problem and they computed non-perturbatively some correlation functions in
the 2D WZ model by employing a lattice formulation
of~Ref.~\cite{Kikukawa:2002as}. They considered the 2D $\mathcal{N}=(2,2)$ WZ
model with a massless cubic superpotential
\begin{equation}
   W(\Phi)=\frac{\lambda}{3}\Phi^3,
\label{eq:(1.1)}
\end{equation}
which, according to the conjectured correspondence, should provide a LG
description of a pair of the $\mathcal{N}=2$ $c=1$ minimal models, where one is
left-moving and the other is right-moving (the so-called $A_2$ model). In the
IR limit, the scalar field in the WZ model~$A$ is identified with a chiral
primary field in the $A_2$~model with the conformal
dimensions~$(h,\Bar{h})=(1/6,1/6)$ and~$U(1)$ charges~$(q,\Bar{q})=(1/3,1/3)$.
(The complex conjugate $A^*$ is identified with an anti-chiral primary field
with $(h,\Bar{h})=(1/6,1/6)$ and~$(q,\Bar{q})=(-1/3,-1/3)$.) The authors
of~Ref.~\cite{Kawai:2010yj} obtained finite-size scalings of scalar two-point
functions which are remarkably consistent with the above SCFT correspondence,
thus demonstrated the power of a lattice formulation of this supersymmetric
field theory.\footnote{For preceding numerical simulations of the 2D
$\mathcal{N}=(2,2)$ WZ model with a \emph{massive} cubic superpotential
$W(\Phi)=m\Phi^2/2+\lambda\Phi^3/3$, see Refs.~\cite{Beccaria:1998vi,%
Catterall:2001fr,Giedt:2005ae,Bergner:2007pu,Kastner:2008zc,Synatschke:2010jn}.
See also Refs.~\cite{Sakai:1983dg,Fujikawa:2002pa} for theoretical background.}

In this paper, motivated by the success of~Ref.~\cite{Kawai:2010yj}, we study
the 2D $\mathcal{N}=(2,2)$ WZ model with massless cubic
superpotential~\eqref{eq:(1.1)} numerically. We employ a non-perturbative
formulation advocated in~Ref.~\cite{Kadoh:2009sp} that uses a simple momentum
cutoff regularization. Although there is an issue concerning the locality in
this formulation, the restoration of an expected locality property can be shown
at least within perturbation theory~\cite{Kadoh:2009sp}. This formulation
possesses very nice symmetry properties: it exactly preserves full SUSY,
translational invariance, and linear internal symmetries such as the
$R$-symmetry. We believe that these nice symmetry properties are especially
useful in defining Noether currents in the regularized framework. In fact, by
defining conserved supercurrents and identifying the component of the
superconformal currents in the IR limit, we numerically measure the central
charge of the system in the IR region: Together with a measurement of the
conformal dimension, this forms a main result of the present paper.

Throughout this paper, Greek indices from the middle of the alphabet, $\mu$,
$\nu$, \dots\ run over $0$ and~$1$. Greek indices from the beginning $\alpha$,
$\beta$, \dots\ are for spinor indices and run over $1$ and~$2$. Repeated
indices are \emph{not} summed over unless explicit summation symbol is
indicated. We extensively use the complex coordinates defined by
\begin{equation}
   z\equiv x_0+ix_1,\qquad\Bar{z}\equiv x_0-ix_1,
\end{equation}
and
\begin{equation}
   \partial_z\equiv\frac{1}{2}(\partial_0-i\partial_1),\qquad
   \partial_{\Bar{z}}\equiv\frac{1}{2}(\partial_0+i\partial_1).
\end{equation}
Conjugate momenta are defined by
\begin{equation}
   p_z\equiv\frac{1}{2}(p_0-ip_1),\qquad
   p_{\Bar{z}}\equiv\frac{1}{2}(p_0+ip_1).
\end{equation}
Two-dimensional gamma matrices are defined by
\begin{equation}
   \gamma_0\equiv\begin{pmatrix}0&1\\1&0\end{pmatrix},\qquad
   \gamma_1\equiv\begin{pmatrix}0&i\\-i&0\end{pmatrix},
\end{equation}
and
\begin{equation}
   \gamma_z\equiv\begin{pmatrix}0&1\\0&0\end{pmatrix},\qquad
   \gamma_{\Bar{z}}\equiv\begin{pmatrix}0&0\\1&0\end{pmatrix}.
\end{equation}

\section{Supersymmetric formulation of the 2D $\mathcal{N}=(2,2)$ WZ
model}
We start by recapitulating the formulation of~Ref.~\cite{Kadoh:2009sp}. We
suppose that the system is defined in a finite box with a physical
size~$L_0\times L_1$. The Fourier modes~$\Tilde{f}(p)$ of a periodic function
in the box are defined by
\begin{equation}
   f(x)=\frac{1}{L_0L_1}\sum_pe^{ipx}\Tilde{f}(p),\qquad
   \Tilde{f}(p)=\int d^2x\,e^{-ipx}f(x),
\end{equation}
where the momentum~$p$ takes discrete values
\begin{equation}
   p_\mu=\frac{2\pi}{L_\mu}\,n_\mu,\qquad n_\mu=0,\pm1,\pm2,\dots,
\end{equation}
and by the definition,
\begin{equation}
   \Tilde{f^*}(p)=\Tilde{f}(-p)^*,
\end{equation}
where the left-hand side denotes the Fourier transformation of the complex
conjugate of~$f(x)$, $f(x)^*$. In the present formulation~\cite{Kadoh:2009sp},
we restrict the momentum~$p$ by an ultraviolet (UV) cutoff~$\Lambda$,
\begin{equation}
   -\Lambda\leq p_\mu\leq\Lambda,\qquad\text{for $\mu=0$, $1$}.
\label{eq:(2.4)}
\end{equation}
We parametrize this UV cutoff by a ``lattice spacing''~$a$,
\begin{equation}
   \Lambda\equiv\frac{\pi}{a},   
\label{eq:(2.5)}
\end{equation}
although we do not assume an underlying spacetime lattice structure in this
paper (see below). Throughout this paper, all dimensionful quantities are
measured in units of the lattice spacing. In particular, if we define the
number of lattice points $N_\mu$ by
\begin{equation}
   L_\mu=N_\mu a,
\end{equation}
then imposing Eq.~\eqref{eq:(2.4)} implies
\begin{equation}
   -\frac{N_\mu}{2}\leq n_\mu\leq\frac{N_\mu}{2},\qquad\text{for $\mu=0$, $1$},
\end{equation}
and thus the number of points in the momentum grid is given
by~$\sum_p1=(N_0+1)(N_1+1)$.

In the present formulation of the 2D $\mathcal{N}=(2,2)$ WZ model, the
partition function is defined by
\begin{equation}
   \mathcal{Z}\equiv
   \int\prod_{-\Lambda\leq p_\mu\leq\Lambda}
   \left[d\Tilde{A}(p)\,d\Tilde{A^*}(p)
   \prod_{\alpha=1}^2d\Tilde{\psi}_\alpha(p)
   \prod_{\Dot{\alpha}=\Dot{1}}^{\Dot{2}}d\Tilde{\Bar{\psi}}_{\Dot{\alpha}}(p)
   \,d\Tilde{F}(p)\,d\Tilde{F^*}(p)\right]e^{-S},
\label{eq:(2.8)}
\end{equation}
where $A$, $(\psi_\alpha,\Bar{\psi}_{\Dot{\alpha}})$, and~$F$ are the scalar,
fermion and auxiliary fields, respectively. In this expression, the action~$S$
is simply the action of the continuum WZ model in terms of Fourier modes:
\begin{align}
   S&=\frac{1}{L_0L_1}\sum_p
   \Biggl[4p_z\Tilde{A^*}(-p)p_{\Bar{z}}\Tilde{A}(p)
   -\Tilde{F^*}(-p)\Tilde{F}(p)
\notag\\
   &\qquad\qquad\qquad{}
   -\Tilde{F^*}(-p)*W'(\Tilde{A})^*(p)
   -\Tilde{F}(-p)*W'(\Tilde{A})(p)
\notag\\
   &\qquad\qquad\qquad{}
   +(\Tilde{\Bar{\psi}}_{\Dot{1}},\Tilde{\psi}_2)(-p)
   \begin{pmatrix}
   2ip_z&W''(\Tilde{A})^**\\
   W''(\Tilde{A})*&2ip_{\Bar{z}}\\
   \end{pmatrix}
   \begin{pmatrix}
   \Tilde{\psi}_1\\
   \Tilde{\Bar{\psi}}_{\Dot{2}}\\
   \end{pmatrix}(p)
   \Biggr],
\label{eq:(2.9)}
\end{align}
where the holomorphic function~$W(A)$ is the superpotential and $*$~denotes
the convolution
\begin{equation}
   \left(\Tilde{\varphi}_1*\Tilde{\varphi}_2\right)(p)\equiv
   \frac{1}{L_0L_1}\sum_q
   \Tilde{\varphi}_1(q)\Tilde{\varphi}_2(p-q).
\end{equation}
Products contained in $W'(\Tilde{A})$ and~$W''(\Tilde{A})$ are also understood
as convolutions.

Since action~\eqref{eq:(2.9)} is identical to that in the continuum theory,
the regularized theory~\eqref{eq:(2.8)} is manifestly invariant under all
symmetries that are consistent with the momentum restriction~\eqref{eq:(2.4)}.
This is the case for symmetry transformations that act linearly on field
variables. Thus, in the present formulation, SUSY, translational invariance,
and the $R$-symmetry (if it exists) are exactly preserved. One can derive
Ward--Takahashi (WT) identities associated with these symmetries in a
regularized framework.

What is sacrificed in the present formulation, on the other hand, is locality.
One sees that the kinetic terms and the interaction terms are quite non-local
in the configuration space when the UV cutoff~$\Lambda$ is finite. In fact,
when both integers $N_0$ and~$N_1$ are odd, the present formulation is nothing
but a two-dimensional version of a lattice formulation of the four-dimensional
WZ model studied in~Ref.~\cite{Bartels:1983wm} that is based on the SLAC
derivative~\cite{Drell:1976bq,Drell:1976mj}. A detailed analysis of a
one-dimensional version (i.e., quantum mechanics) can be found
in~Ref.~\cite{Bergner:2009vg}. Thus there is an issue of whether the present
formulation reproduces an IR physics expected in the original target theory
(i.e., whether it belongs to the same universality class or not). Although one
can show~\cite{Kadoh:2009sp} that within perturbation theory the expected
locality is restored in the limit~$\Lambda\to\infty$ for two- and
three-dimensional WZ models, the validity of the present formulation at the
non-perturbative level is not obvious \textit{a priori}. We believe that our
results in this paper will provide an affirmative indication regarding this
question of locality.

As a side remark, we note that the present formulation cannot be regarded as a
spacetime lattice formulation when either $N_0$ or~$N_1$ is even. Fourier
modes~$\Tilde{f}(p)$ of a function~$f(x)$ defined on a spacetime lattice is
periodic in the Brillouin zone $\Tilde{f}(p+(\pi/a)\Hat{\mu})=\Tilde{f}(p)$,
where $\Hat{\mu}$ denotes a unit vector in the $\mu$ direction. However, the
combination~$p_\mu\Tilde{f}(p)$ appearing in the action breaks this periodicity
at the boundary of the Brillouin zone and this cannot be regarded as a Fourier
transformation of a function defined on a lattice. (When $N_\mu$ are odd, there
is no Fourier mode on the boundary of the Brillouin zone.) Throughout this
paper we set $N_\mu$ even and this means that we lose a connection with a
spacetime lattice formulation; we have to regard our configuration space as
continuous (with a limited resolution). Still, Eq.~\eqref{eq:(2.8)} provides a
regularized partition function and can serve as a starting point for
non-perturbative study.

\section{Simulation algorithm based on the Nicolai map}

After integrating over the auxiliary fields $\Tilde{F}$ and~$\Tilde{F^*}$
in~Eq.~\eqref{eq:(2.8)}, the partition function becomes
\begin{equation}
   \mathcal{Z}\equiv
   \int\prod_{-\Lambda\leq p_\mu\leq\Lambda}
   \left[d\Tilde{A}(p)\,d\Tilde{A^*}(p)
   \prod_{\alpha=1}^2d\Tilde{\psi}_\alpha(p)
   \prod_{\Dot{\alpha}=\Dot{1}}^{\Dot{2}}
   d\Tilde{\Bar{\psi}}_{\dot\alpha}(p)
   \right]e^{-S},
\end{equation}
where
\begin{align}
   S&=\frac{1}{L_0L_1}\sum_p
   \biggl[\Tilde{N^*}(-p)\Tilde{N}(p)
\notag\\
   &\qquad\qquad\qquad{}+
   (\Tilde{\Bar{\psi}}_{\dot1},\Tilde{\psi}_2)(-p)
   \begin{pmatrix}
     2ip_z&W''(\Tilde{A})^**\\
     W''(\Tilde{A})*&2ip_{\Bar{z}}\\
   \end{pmatrix}
   \begin{pmatrix}
     \Tilde{\psi}_1\\
     \Tilde{\Bar{\psi}}_{\dot2}\\
   \end{pmatrix}(p)
   \Biggr].
\end{align}
In this expression, $\Tilde{N}(p)$ is a function that specifies the Nicolai
map~\cite{Nicolai:1979nr,Nicolai:1980jc,Cecotti:1981fu,Parisi:1982ud,%
Cecotti:1982ad},
\begin{align}
   &\Tilde{N}(p)\equiv2ip_z\Tilde{A}(p)+W'(\Tilde{A})^*(p),
\\
   &\Tilde{N^*}(-p)=\Tilde{N}(p)^*
   =-2ip_{\Bar{z}}\Tilde{A^*}(-p)+W'(\Tilde{A})(-p).
\end{align}
For example, for cubic superpotential~\eqref{eq:(1.1)}, the explicit form of
the function~$\Tilde{N}(p)$ is given by
\begin{align}
   \Tilde{N}(p)&=
    i(p_0-ip_1)\Tilde{A}(p)
   +\lambda\frac{1}{L_0L_1}\sum_q\Tilde{A^*}(q)\Tilde{A^*}(p-q)
\notag\\
   &=i(p_0-ip_1)\Tilde{A}(p)
   +\lambda\frac{1}{L_0L_1}\sum_q\Tilde{A}(q)^*\Tilde{A}(-p-q)^*.
\end{align}
We then note
\begin{equation}
   S=\frac{1}{L_0L_1}\sum_p
   \left[\Tilde{N^*}(-p)\Tilde{N}(p)
   +(\Tilde{\Bar{\psi}}_{\dot1},\Tilde{\psi}_2)(-p)
   \begin{pmatrix}
   \frac{\partial\Tilde{N}(p)}{\partial\Tilde{A}(p)}
   &\frac{\partial\Tilde{N}(p)}{\partial\Tilde{A^*}(p)}*\\
   \frac{\partial\Tilde{N^*}(p)}{\partial\Tilde{A}(p)}*
   &\frac{\partial\Tilde{N^*}(p)}{\partial\Tilde{A^*}(p)}\\
   \end{pmatrix}
   \begin{pmatrix}
     \Tilde{\psi}_1\\
     \Tilde{\Bar{\psi}}_{\dot2}\\
   \end{pmatrix}(p)
   \right],
\end{equation}
and therefore, after integrating over fermion fields,
\begin{equation}
   \mathcal{Z}\equiv
   \int\prod_{-\Lambda\leq p_\mu\leq\Lambda}
   \left[d\Tilde{A}(p)\,d\Tilde{A^*}(p)\right]
   \exp\left[\frac{1}{L_0L_1}\sum_p\Tilde{N^*}(-p)\Tilde{N}(p)\right]
   \det\frac{\partial(\Tilde{N},\Tilde{N^*})}{\partial(\Tilde{A},\Tilde{A^*})}.
\end{equation}
In this partition function, we may change integration variables from
$(\Tilde{A}(p),\Tilde{A^*}(p))$ to~$(\Tilde{N}(p),\Tilde{N^*}(p))$. Then the
Jacobian associated with this change of variables precisely cancels the
absolute value of the fermion determinant. In this way, we arrive at the
expression
\begin{align}
   \mathcal{Z}&=
   \int\prod_{-\Lambda\leq p_\mu\leq\Lambda}
   \left[d\Tilde{N}(p)\,d\Tilde{N^*}(p)\right]
   \exp\left[-\frac{1}{L_0L_1}\sum_p\Tilde{N}(p)^*\Tilde{N}(p)\right]
\notag\\
   &\qquad\qquad\qquad\qquad\qquad\qquad{}
   \times
   \sum_i\left.\sign\det
   \frac{\partial(\Tilde{N},\Tilde{N^*})}
   {\partial(\Tilde{A},\Tilde{A^*})}\right|%
   _{\Tilde{A}=\Tilde{A}_i,\Tilde{A^*}=\Tilde{A^*}_i},
\label{eq:(3.8)}
\end{align}
where $\Tilde{A}(p)_i$ and~$\Tilde{A^*}(p)_i$ ($i=1$, $2$, \dots) are
solutions of
\begin{align}
   2ip_z\Tilde{A}(p)+W'(\Tilde{A})(-p)^*-\Tilde{N}(p)&=0,
\label{eq:(3.9)}
\\
   -2ip_{\Bar{z}}\Tilde{A}(p)^*+W'(\Tilde{A})(-p)-\Tilde{N}(p)^*&=0.
\label{eq:(3.10)}
\end{align}

Representation~\eqref{eq:(3.8)} for the partition function (and a similar
representation for expectation values of observables) gives rise to the
following simulation algorithm~\cite{Beccaria:1998vi,Kawai:2010yj}:
\begin{itemize}
\item[(i)] Generate Gaussian random numbers $(\Tilde{N}(p),\Tilde{N}(p)^*)$.
\item[(ii)] Find (numerically) all the solutions
$(\Tilde{A}(p)_i,\Tilde{A}(p)^*_i)$ ($i=1$, $2$, \dots)
of~Eqs.~\eqref{eq:(3.9)} and~\eqref{eq:(3.10)}.
\item[(iii)] Then compute the sums
\begin{equation}
   \sum_i\left.\sign\det
   \frac{\partial(\Tilde{N},\Tilde{N^*})}
   {\partial(\Tilde{A},\Tilde{A^*})}\right|%
   _{\Tilde{A}=\Tilde{A}_i,\Tilde{A^*}=\Tilde{A^*}_i},
\end{equation}
and
\begin{equation}
   \sum_i\left.
   \sign\det
   \frac{\partial(\Tilde{N},\Tilde{N^*})}
   {\partial(\Tilde{A},\Tilde{A^*})}\,
   \mathcal{O}(\Tilde{A},\Tilde{A^*})
   \right|%
   _{\Tilde{A}=\Tilde{A}_i,\Tilde{A^*}=\Tilde{A^*}_i},
\end{equation}
where $\mathcal{O}(\Tilde{A},\Tilde{A^*})$ is an observable.
\item[(iv)] Repeat the steps from~(i) and take an average over the Gaussian
random numbers. The expectation value of an observable~$\mathcal{O}$ is given
by
\begin{equation}
   \left\langle\mathcal{O}\right\rangle
   =\frac{1}{\Delta}
   \frac{\left\langle
   \sum_i\left.
   \sign\det
   \frac{\partial(\Tilde{N},\Tilde{N^*})}
   {\partial(\Tilde{A},\Tilde{A^*})}\,
   \mathcal{O}(\Tilde{A},\Tilde{A^*})
   \right|%
   _{\Tilde{A}=\Tilde{A}_i,\Tilde{A^*}=\Tilde{A^*}_i}\right\rangle_N}
   {\left\langle1\right\rangle_N},
\label{eq:(3.13)}
\end{equation}
where $\langle\cdot\rangle_N$ denotes the average over the Gaussian random
numbers and
\begin{equation}
   \Delta\equiv
   \frac{\left\langle
   \sum_i\left.
   \sign\det
   \frac{\partial(\Tilde{N},\Tilde{N^*})}
   {\partial(\Tilde{A},\Tilde{A^*})}\right|%
   _{\Tilde{A}=\Tilde{A}_i,\Tilde{A^*}=\Tilde{A^*}_i}\right\rangle_N}
   {\left\langle1\right\rangle_N}.
\label{eq:(3.14)}
\end{equation}
\end{itemize}
As Eq.~\eqref{eq:(3.13)} shows, in the present algorithm the expectation value
of an observable~$\langle\mathcal{O}\rangle$ is given by the ratio of two
expectation values with respect to the Gaussian random numbers. In this paper,
we apply a simple propagation of error rule to the ratio to estimate the
statistical error in~$\langle\mathcal{O}\rangle$. This procedure must be good
enough because the statistical error in the denominator of the ratio, $\Delta$,
is generally quite small (see below).

Since the above simulation algorithm is based on the generation of the Gaussian
random numbers, there is no autocorrelation among generated configurations and
there is no critical slowing down; this is an overwhelming advantage of the
present simulation algorithm.\footnote{We would like thank Martin L\"uscher for
bringing our attention to this point.} Another interesting feature of this
algorithm is that the ``normalized partition function''
$\Delta$ in~Eq.~\eqref{eq:(3.14)} can be argued to give the Witten index
$\tr(-1)^F$~\cite{Witten:1982df}.\footnote{%
With conventional simulation algorithms, one needs more elaborate method to
compute the Witten index; see~Ref.~\cite{Kanamori:2010gw}.}
For the massive free theory $W(\Phi)=m\Phi^2/2$, one easily sees that
$\Delta=1$ and reproduces the correct Witten index~$\tr(-1)^F=1$. Thus assuming
that the proportionality constant between the partition function defined by the
functional integral and~$\tr(-1)^F$ is independent of the interaction, we have
$\Delta=\tr(-1)^F$. It is known that for a homogeneous superpotential
$W(\Phi)=\lambda\Phi^n/n$, $\tr(-1)^F=n-1$. Therefore, the
relation~$\Delta=n-1$ provides a quantitative test of the
simulation~\cite{Kawai:2010yj}.

A disadvantage of the present simulation algorithm is that it is not
\textit{a priori} clear how many solutions $(\Tilde{A}(p)_i,\Tilde{A}(p)^*_i)$
that Eqs.~\eqref{eq:(3.9)} and~\eqref{eq:(3.10)} have. Thus we cannot be
completely sure whether we have found all the solutions or not. Another
limitation of the algorithm is that it is applicable only to supersymmetric
boundary conditions; so it cannot explore physics with the finite temperature,
for example.

\section{Simulation parameters}
In this paper, we fix the coupling constant in~Eq.~\eqref{eq:(1.1)} to be
\begin{equation}
   a\lambda=0.3.
\end{equation}
This is the same choice of the coupling constant as~Ref.~\cite{Kawai:2010yj},
provided that our parameter~$a$ in~Eq.~\eqref{eq:(2.5)} is identified with the lattice spacing
in~Ref.~\cite{Kawai:2010yj}. The size of the momentum grid $N_0\times N_1$ is
varied from~$16\times16$ to~$36\times36$. For each value of~$N_0\times N_1$, we
generated $1280$ configurations of the Gaussian random
numbers~$(\Tilde{N}(p),\Tilde{N}(p)^*)$. We then solved Eqs.~\eqref{eq:(3.9)}
and~\eqref{eq:(3.10)} by using the Newton--Raphson (NR) method~\cite{NR3}. The
solution $(\Tilde{A}(p),\Tilde{A}(p)^*)$ depends on the initial guess in the
NR method and, as the initial guess, we used $100$ random
configurations~$(\Tilde{A}(p),\Tilde{A}(p)^*)$ generated with Gaussian random
numbers with a unit variant. We judged that the convergence of the NR method is
achieved when the maximum norm of the residue of~Eqs.~\eqref{eq:(3.9)}
and~\eqref{eq:(3.10)} becomes smaller than~$10^{-13}$. Two obtained solutions
were regarded identical when the the maximum norm of the difference of the
solutions is smaller than~$10^{-11}$. In this way, we obtained configurations
tabulated in~Tables~\ref{table:1} and~\ref{table:2}. The total amount of
computational time was $34\,307.7\,\text{core}\cdot\text{hour}$
%$8576.9\,\text{CPU}\cdot\text{hour}$
on the Intel Xeon $2.93\text{GHz}$.
\begin{table}
\caption{Classification of obtained configurations. $\Delta$ is given
by~Eq.~\eqref{eq:(3.14)} and $\delta$ is defined by~Eq.~\eqref{eq:(5.8)}.}
\label{table:1}
\begin{center}
%\catcode`?=\active\def?{\phantom{0}}
\begin{tabular}{lrrrrrr}
\hline
$N_0=N_1$
& $16$ & $18$ & $20$ & $22$ & $24$ & $26$\\
\hline
$(+,+)$
& $1276$ & $1273$ & $1275$ & $1271$ & $1271$ & $1273$\\
$(-,+,+,+)$
& $4$ & $7$ & $5$ & $9$ & $9$ & $7$\\
$(+)$
& $0$ & $0$ & $0$ & $0$ & $0$ & $0$\\
$(+,+,+)$
& $0$ & $0$ & $0$ & $0$ & $0$ & $0$\\
$(-,+,+,+,+)$
& $0$ & $0$ & $0$ & $0$ & $0$ & $0$\\
$(-,-,+,+,+,+)$
& $0$ & $0$ & $0$ & $0$ & $0$ & $0$\\
\hline
$\Delta$
& $2$ & $2$ & $2$ & $2$ & $2$ & $2$\\
\hline
$\delta$ [\%]
& $0.2(2)$ & $0.1(1)$ & ${-}0.2(1)$ & $0.1(1)$ & ${-}0.1(1)$ & ${-}0.1(1)$\\
\hline
\end{tabular}
\end{center}
\end{table}
\begin{table}
\caption{Classification of obtained configurations (continued). $\Delta$ is
given by~Eq.~\eqref{eq:(3.14)} and $\delta$ is defined
by~Eq.~\eqref{eq:(5.8)}.}
\label{table:2}
\begin{center}
\begin{tabular}{lrrrrr}
\hline
$N_0=N_1$
& $28$ & $30$ & $32$ & $34$ & $36$\\
\hline
$(+,+)$
& $1264$ & $1261$ & $1250$ & $1254$ & $1221$\\
$(-,+,+,+)$
& $16$ & $17$ & $24$ & $17$ & $26$\\
$(+)$
& $0$ & $1$ & $2$ & $6$ & $31$\\
$(+,+,+)$
& $0$ & $0$ & $4$ & $2$ & $2$\\
$(-,+,+,+,+)$
& $0$ & $1$ & $0$ & $0$ & $0$\\
$(-,-,+,+,+,+)$
& $0$ & $0$ & $0$ & $1$ & $0$\\
\hline
$\Delta$
& $2$ & $2.000(1)$ & $2.002(2)$ & $1.997(2)$ & $1.977(4)$ \\
\hline
$\delta$ [\%]
& ${-}0.0(1)$ & $0.1(2)$ & $0.0(2)$ & ${-}0.1(3)$ & $0.0(5)$\\
\hline
\end{tabular}
\end{center}
\end{table}
In~Tables~\ref{table:1} and~\ref{table:2}, configurations are classified
according to the number of associated solutions
$(\Tilde{A}(p)_i,\Tilde{A}(p)^*_i)$ and the associated sign of the
Jacobian~\cite{Kawai:2010yj}. For example, an entry in a row with the
symbol~$(-,+,+,+)$ implies that there was a configuration
of~$(\Tilde{N}(p),\Tilde{N}(p)^*)$ for which we have four different solutions
$(\Tilde{A}(p)_i,\Tilde{A}(p)^*_i)$ ($i=1$, $2$, $3$, $4$) and the sign of the
Jacobian is negative for one of them and it is positive for other three. Thus,
we can read off $\Delta$ in~Eq.~\eqref{eq:(3.14)} from this table. We see that
the expected relation $\Delta=2$ for cubic superpotential~\eqref{eq:(1.1)}
actually holds almost within one percent even for the worst case, for
$N_0\times N_1=36\times36$. This fact suggests that even if our root-finding
code missed some of solutions of~Eqs.~\eqref{eq:(3.9)} and~\eqref{eq:(3.10)},
they are precious few.

\section{SUSY WT identities}
WT identities associated with exact symmetries in the formulation provide a
consistency check for the numerical simulation. In particular, since SUSY is an
exact symmetry in the present formulation, SUSY WT identities must hold even
with a \emph{finite} UV cutoff. This aspect is quite different from
conventional lattice formulations in which one hopes that SUSY is restored only
in the continuum limit. Thus an explicit confirmation of SUSY WT identities
in the present formulation is interesting in its own right.

SUSY transformations in the 2D $\mathcal{N}=(2,2)$ WZ model are given by $Q_1$,
$Q_2$, $Q_{\Dot{1}}$, and~$Q_{\Dot{2}}$. In terms of Fourier modes, they are given
by
\begin{align}
   &Q_1\Tilde{\Bar{\psi}}_{\Dot{1}}(p)
   =-2\sqrt{2}ip_{\Bar{z}}\Tilde{A^*}(p),
   &
   &Q_1\Tilde{A^*}(p)=0,
\notag\\
   &Q_1\Tilde{F^*}(p)
   =2\sqrt{2}ip_{\Bar{z}}\Tilde{\Bar{\psi}}_{\Dot{2}}(p),
   &
   &Q_1\Tilde{\Bar{\psi}}_{\Dot{2}}(p)=0,
\notag\\
   &Q_1\Tilde{A}(p)
   =\sqrt{2}\Tilde{\psi}_1(p),
   &
   &Q_1\Tilde{\psi}_1(p)=0,
\notag\\
   &Q_1\Tilde{\psi}_2(p)
   =\sqrt{2}\Tilde{F}(p),
   &
   &Q_1\Tilde{F}(p)=0,
\label{eq:(5.1)}
\end{align}
for~$Q_1$,
\begin{align}
   &Q_2\Tilde{\Bar{\psi}}_{\Dot{2}}(p)
   =-2\sqrt{2}ip_z\Tilde{A^*}(p),
   &
   &Q_2\Tilde{A^*}(p)=0,
\notag\\
   &Q_2\Tilde{F^*}(p)
   =-2\sqrt{2}ip_z\Tilde{\Bar{\psi}}_{\Dot{1}}(p),
   &
   &Q_2\Tilde{\Bar{\psi}}_{\Dot{1}}(p)=0,
\notag\\
   &Q_2\Tilde{A}(p)
   =\sqrt{2}\Tilde{\psi}_2(p),
   &
   &Q_2\Tilde{\psi}_2(p)=0,
\notag\\
   &Q_2\Tilde{\psi}_1(p)
   =-\sqrt{2}\Tilde{F}(p),
   &
   &Q_2\Tilde{F}(p)=0,
\label{eq:(5.2)}
\end{align}
for~$Q_2$
\begin{align}
   &\Bar{Q}_{\Dot{1}}\Tilde{\psi}_1(p)
   =-2\sqrt{2}ip_{\Bar{z}}\Tilde{A}(p),
   &
   &\Bar{Q}_{\Dot{1}}\Tilde{A}(p)=0,
\notag\\
   &\Bar{Q}_{\Dot{1}}\Tilde{F}(p)
   =-2\sqrt{2}ip_{\Bar{z}}\Tilde{\psi}_2(p),
   &
   &\Bar{Q}_{\Dot{1}}\Tilde{\psi}_2(p)=0,
\notag\\
   &\Bar{Q}_{\Dot{1}}\Tilde{A^*}(p)
   =\sqrt{2}\Tilde{\Bar{\psi}}_{\Dot{1}}(p),
   &
   &\Bar{Q}_{\Dot{1}}\Tilde{\Bar{\psi}}_{\Dot{1}}(p)=0,
\notag\\
   &\Bar{Q}_{\Dot{1}}\Tilde{\Bar{\psi}}_{\Dot{2}}(p)
   =-\sqrt{2}\Tilde{F^*}(p),
   &
   &\Bar{Q}_{\Dot{1}}\Tilde{F^*}(p)=0,
\label{eq:(5.3)}
\end{align}
for~$\Bar{Q}_{\Dot{1}}$,
\begin{align}
   &\Bar{Q}_{\Dot{2}}\Tilde{\psi}_2(p)
   =-2\sqrt{2}ip_z\Tilde{A}(p),
   &
   &\Bar{Q}_{\Dot{2}}\Tilde{A}(p)=0,
\notag\\
   &\Bar{Q}_{\Dot{2}}\Tilde{F}(p)
   =2\sqrt{2}ip_z\Tilde{\psi}_1(p),
   &
   &\Bar{Q}_{\Dot{2}}\Tilde{\psi}_1(p)=0,
\notag\\
   &\Bar{Q}_{\Dot{2}}\Tilde{A^*}(p)
   =\sqrt{2}\Tilde{\Bar{\psi}}_{\Dot{2}}(p),
   &
   &\Bar{Q}_{\Dot{2}}\Tilde{\Bar{\psi}}_{\Dot{2}}(p)=0,
\notag\\
   &\Bar{Q}_{\Dot{2}}\Tilde{\Bar{\psi}}_{\Dot{1}}(p)
   =\sqrt{2}\Tilde{F^*}(p),
   &
   &\Bar{Q}_{\Dot{2}}\Tilde{F^*}(p)=0,
\label{eq:(5.4)}
\end{align}
for~$\Bar{Q}_{\Dot{2}}$.

A simple one-point SUSY WT identity is given by~\cite{Catterall:2001fr}
\begin{equation}
   0=\left\langle S\right\rangle
   =\left\langle S_B\right\rangle
   +\left(\sum_p1-2\sum_p1\right)
   =\left\langle S_B\right\rangle-(N_0+1)(N_1+1),
\label{eq:(5.5)}
\end{equation}
where $S_B$ is the bosonic part of the action after integrating over the
auxiliary field, giving by
\begin{equation}
   S_B\equiv\frac{1}{L_0L_1}\sum_p\Tilde{N^*}(-p)\Tilde{N}(p).
\end{equation}
In Eq.~\eqref{eq:(5.5)}, the first equality holds because the action~$S$
in~Eq.~\eqref{eq:(2.9)} can be written as
\begin{align}
   S&=-\frac{1}{2}Q\frac{1}{L_0L_1}\sum_p
   \Bigl\{
   \Bigl[2ip_z\Tilde{A^*}(-p)+\Tilde{F^*}(-p)+2W'(\Tilde{A})*
   \Bigr]\Tilde{\psi}_1(p)
\notag\\
   &\qquad\qquad\qquad\qquad{}
   +\Bigl[-2ip_{\Bar{z}}\Tilde{A}(p)+\Tilde{F}(p)+2W'(\Tilde{A})^**
   \Bigr]\Tilde{\Bar{\psi}}_{\Dot{2}}(-p)
   \Bigr\},
\end{align}
where $Q\equiv-(\Bar{Q}_{\Dot{1}}+Q_2)/\sqrt{2}$ is a nilpotent SUSY
transformation. Therefore $\langle S\rangle=0$ provided SUSY is not
spontaneously broken, as the non-zero Witten index in the present system shows.
One can then exactly evaluate the expectation value of the parts of the action
which are quadratic in the auxiliary field and the fermion field; this gives
the second equality in~Eq.~\eqref{eq:(5.5)}. Since $\sum_p1=(N_0+1)(N_1+1)$, we
have the last equality in~Eq.~\eqref{eq:(5.5)}. That is, we have
\begin{equation}
   \delta\equiv\frac{\left\langle S_B\right\rangle}{(N_0+1)(N_1+1)}-1=0.
\label{eq:(5.8)}
\end{equation}
This relation provides a quantitative test of the
simulation~\cite{Catterall:2001fr,Kawai:2010yj}.\footnote{Lattice formulations
adopted in~Refs.~\cite{Catterall:2001fr} and~\cite{Kawai:2010yj} and more
generally lattice formulations based on the Nicolai map~\cite{Sakai:1983dg,%
Beccaria:1998vi,Catterall:2001fr,Kikukawa:2002as,Fujikawa:2002pa,Giedt:2005ae,%
Bergner:2007pu,Kastner:2008zc} possess exact invariance under the above
nilpotent $Q$ transformation. See also Ref.~\cite{Kadoh:2010ca} and references
therein.} We show $\delta$ in~Tables~\ref{table:1} and~\ref{table:2} for each
value of~$N_0\times N_1$ and see that relation~\eqref{eq:(5.8)} holds within
0.5\%.

\begin{figure}[htb]
\centering
\includegraphics[width=120mm]{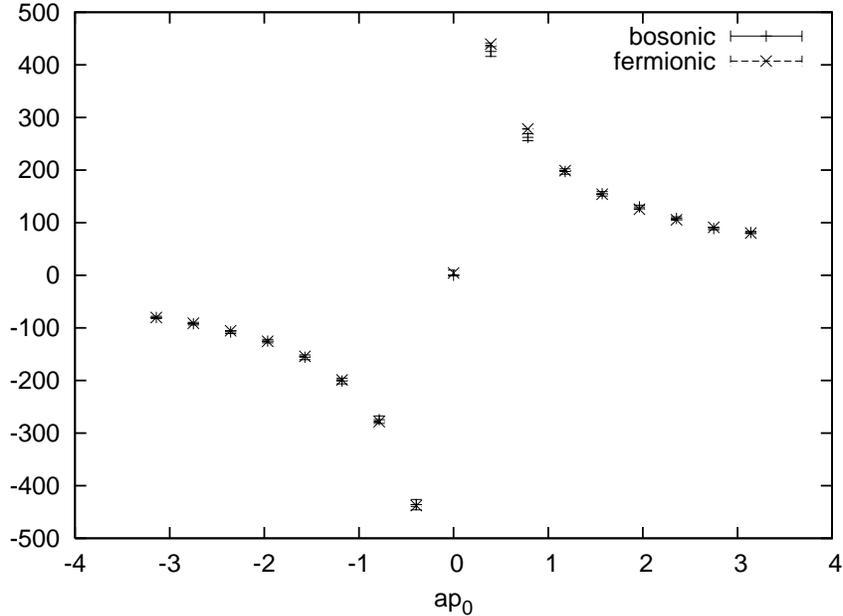}
\caption{The left-hand side (``bosonic'') and the right-hand side
(``fermionic'') of~Eq.~\eqref{eq:(5.9)} as a function of~$ap_0$ along the line
$ap_1=0$; $N_0\times N_1=16\times16$.}
\label{fig:1}
\end{figure}

As for a two-point SUSY WT identity, from the relation
$\langle Q_1(\Tilde{A}(p)\Tilde{\Bar{\psi}}_{\Dot{1}}(-p))\rangle=0$, we
consider
\begin{equation}
   p_0\left\langle\Tilde{A}(p)\Tilde{A^*}(-p)\right\rangle
   =-\im
   \left\langle\Tilde{\psi}_1(p)\Tilde{\Bar{\psi}}_{\Dot{1}}(-p)\right\rangle.
\label{eq:(5.9)}
\end{equation}
In Fig.~\ref{fig:1}, we plotted both sides of this relation for our coarsest
grid ($N_0\times N_1=16\times16$) as a function of~$ap_0$; both sides coincide
within the statistical error.
\begin{figure}[htb]
\centering
\includegraphics[width=120mm]{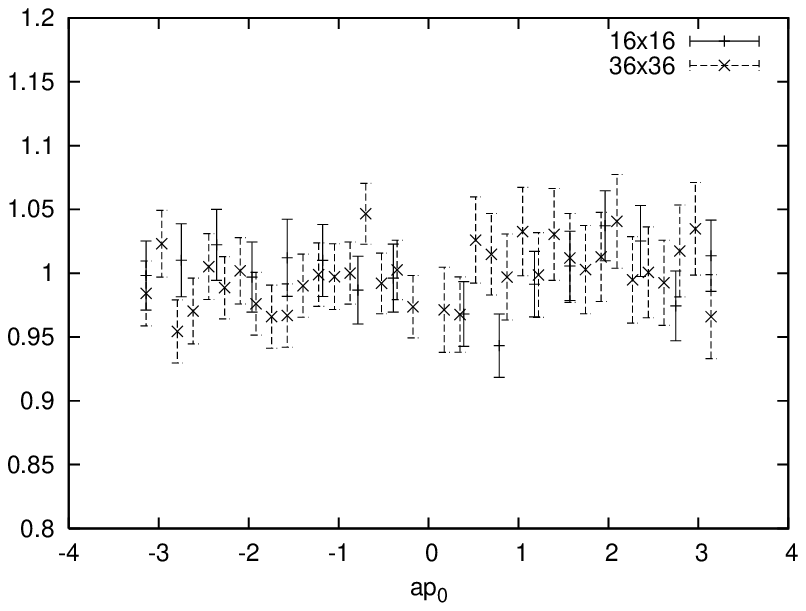}
\caption{The ratio of the left-hand side and the right-hand side of
Eq.~\eqref{eq:(5.9)} as a function of~$ap_0$ along the line $ap_1=0$ (the
origin~$p=0$ is excluded from the plot). The cases
of~$N_0\times N_1=16\times16$ and~$N_0\times N_1=36\times36$ are plotted.}
\label{fig:2}
\end{figure}
For more quantitative comparison, we plotted in~Fig.~\ref{fig:2} the ratio
of the left-hand side and the right-hand side again as a function of~$ap_0$.
(The error in the ratio was estimated by a simple propagation of error rule.)
We plotted the cases of~$N_0\times N_1=16\times16$
and~$N_0\times N_1=36\times36$. For both cases of the momentum grid, the
identity holds within $\sim5\%$, a good indication of SUSY.

\begin{figure}[htb]
\centering
\includegraphics[width=120mm]{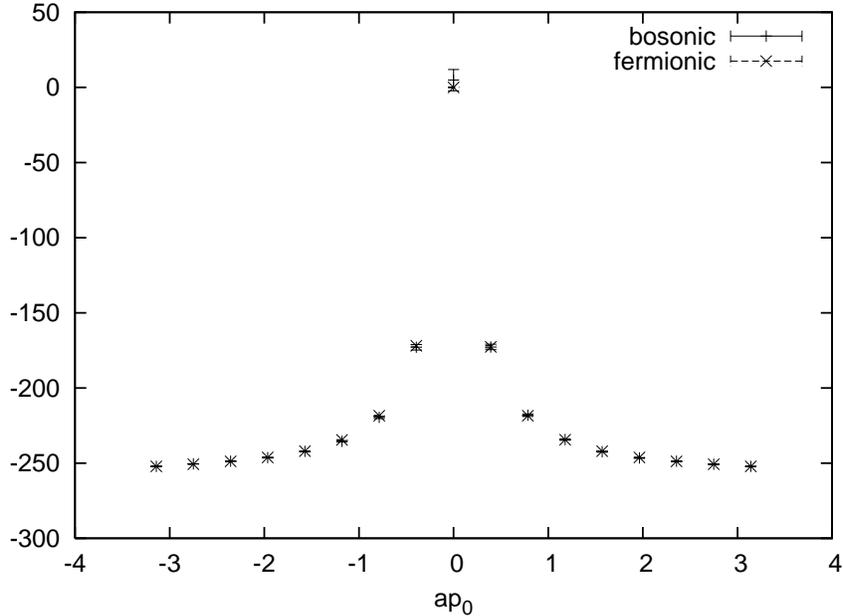}
\caption{The left-hand side (``bosonic'') and the right-hand side
(``fermionic'') of~Eq.~\eqref{eq:(5.10)} along the line $ap_1=0$;
$N_0\times N_1=16\times16$.}
\label{fig:3}
\end{figure}

\begin{figure}[htb]
\centering
\includegraphics[width=120mm]{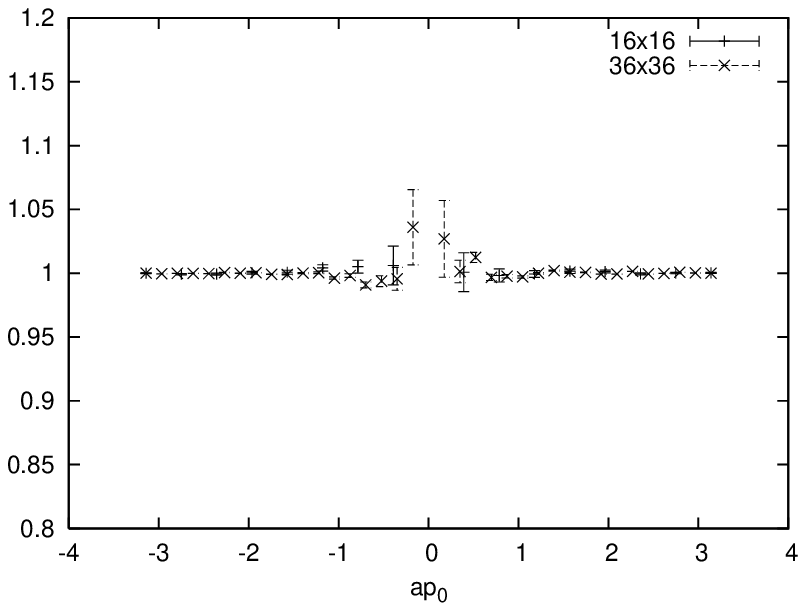}
\caption{The ratio of the left-hand side and the right-hand side of
Eq.~\eqref{eq:(5.10)} as a function of~$ap_0$ along the line $ap_1=0$ (the
origin~$p=0$ is excluded from the plot). The cases
of~$N_0\times N_1=16\times16$ and~$N_0\times N_1=36\times36$ are plotted.}
\label{fig:4}
\end{figure}

Similarly, from $\langle Q_2(\Tilde{F^*}(p)\Tilde{\psi}_1(-p))\rangle=0$,
we have\footnote{%
In the Wess--Zumino model, the integration over the auxiliary field is defined
by an analytic continuation, and according to this prescription, the
correlation function is expressed as
\begin{equation}
   \left\langle\Tilde{F}(p)\Tilde{F^*}(-p))\right\rangle
   =\left\langle
   \left[-\Tilde{N}(p)+2ip_z\Tilde{A}(p)\right]
   \left[-\Tilde{N^*}(-p)-2ip_{\Bar{z}}\Tilde{A^*}(-p)\right]
   \right\rangle-L_0L_1,
\end{equation}
in terms of a correlation function of the Gaussian random fields and the scalar
fields. The last term gives a negative contribution, and despite its
appearance, the left-hand side can be negative as our plot shows.}
\begin{equation}
   \left\langle\Tilde{F}(p)\Tilde{F^*}(-p)\right\rangle
   =-p_1
   \re\left\langle\Tilde{\psi}_1(p)\Tilde{\Bar{\psi}}_{\Dot{1}}(-p)\right\rangle
   +p_0
   \im\left\langle\Tilde{\psi}_1(p)\Tilde{\Bar{\psi}}_{\Dot{1}}(-p)
   \right\rangle.
\label{eq:(5.10)}
\end{equation}
The results for this identity are depicted in~Figs.~\ref{fig:3}
and~\ref{fig:4}.\footnote{Since
$\langle\Tilde{F}(0)\Tilde{F^*}(0)\rangle=
-\langle Q_2(\Tilde{\psi}_1(0)\Tilde{F^*}(0))\rangle/\sqrt{2}$,
$\langle\Tilde{F}(p)\Tilde{F^*}(-p)\rangle$ evaluated at~$p=0$ must vanish if
SUSY is not spontaneously broken. Our plot in~Fig.~\ref{fig:3} is consistent
with this expectation.}

Having obtained these encouraging results concerning SUSY in our numerical
simulation, let us proceed to study physical questions.

\section{Conformal dimension of a chiral primary field}
\begin{figure}[htb]
\centering
\includegraphics[width=120mm]{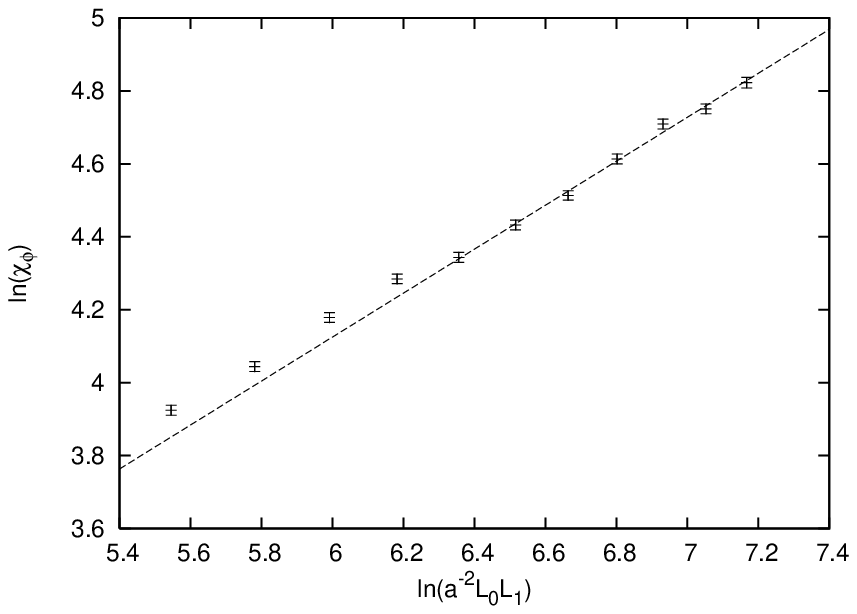}
\caption{$\ln(\chi_\phi)$ as a function of $\ln(a^{-2}L_0L_1)$. No UV
subtraction is made. The broken line is a linear fit with~$1-h-\Bar{h}=0.603$.}
\label{fig:5}
\end{figure}

As noted in the Introduction, in the IR limit, the scalar field~$A$ in the WZ
model with cubic potential~\eqref{eq:(1.1)} is expected to behave as a chiral
primary field with the conformal dimensions~$(h,\Bar{h})=(1/6,1/6)$. For such
a field, the two-point function will behave as
\begin{equation}
   \left\langle A(x)A^*(0)\right\rangle\propto
   \frac{1}{z^{2h}\Bar{z}^{2\Bar{h}}},
   \qquad\text{for $|x|$ large}.
\label{eq:(6.1)}
\end{equation}
Then assuming that $A$~is spinless, $h=\Bar{h}$, the anomalous dimension
$h+\Bar{h}=2h$ could be extracted from the susceptibility of the scalar
field~$\chi_\phi$, defined by~\cite{Kawai:2010yj}
\begin{equation}
   \chi_\phi\equiv
   \frac{1}{a^2}\int_{L_0L_1}d^2x\left\langle A(x)A^*(0)\right\rangle
   =\frac{1}{a^2L_0L_1}\left\langle\left|\Tilde{A}(0)\right|^2\right\rangle,
\label{eq:(6.2)}
\end{equation}
as
\begin{equation}
   \chi_\phi\propto
   \int_{L_0L_1}d^2x\,\frac{1}{(x^2)^{2h}}
   \propto(L_0L_1)^{1-h-\Bar{h}},\qquad\text{for $L_\mu$ large}.
\label{eq:(6.3)}
\end{equation}
Thus $\ln(\chi_\phi)$ would be a linear function of~$\ln(L_0L_1)$ for
large~$L_\mu$ and $1-h-\Bar{h}$ is obtained by the slope of the linear
function~\cite{Kawai:2010yj}. In~Fig.~\ref{fig:5}, we plot $\ln(\chi_\phi)$ as
a function of $\ln(a^{-2}L_0L_1)$ (recall that in our present simulation the
lattice spacing is fixed to $a\lambda=0.3$). From the plot, we see that a fit
by a linear function would be good for large sizes
as~$a^{-2}L_0L_1\gtrsim24\times24$.
Therefore, we applied a linear $\chi^2$ fit by using data
from~$N_0\times N_1=24\times24$ to~$N_0\times N_1=36\times36$. To estimate a
systematic error associated with this choice of fitting range, we also carried
out a linear fit using data from~$N_0\times N_1=26\times26$
to~$N_0\times N_1=36\times36$. See Table~\ref{table:3}
\begin{table}
\caption{Linear $\chi^2$ fit used for a determination of the anomalous
dimension in~Eq.~\eqref{eq:(6.5)}. $n_0\times n_1$ denotes the UV subtraction
region defined in~Eq.~\eqref{eq:(6.4)}.}
\label{table:3}
\begin{center}
\begin{tabular}{llll}
\hline\hline
$n_0\times n_1$ & Fitting range of~$N_0\times N_1$ & $\chi^2/\text{d.o.f.}$ &
$1-h-\Bar{h}$\\
\hline
$0\times0$
& from $24\times24$ to $36\times36$ & $0.904$ & $0.603(19)$\\
$0\times0$
& from $26\times26$ to $36\times36$ & $1.088$ & $0.609(25)$\\
$3\times3$
& from $24\times24$ to $36\times36$ & $0.910$ & $0.624(20)$\\
$3\times3$
& from $26\times26$ to $36\times36$ & $1.108$ & $0.629(26)$\\
\hline\hline
\end{tabular}
\end{center}
\end{table}

\begin{figure}[htb]
\centering
\includegraphics[width=120mm]{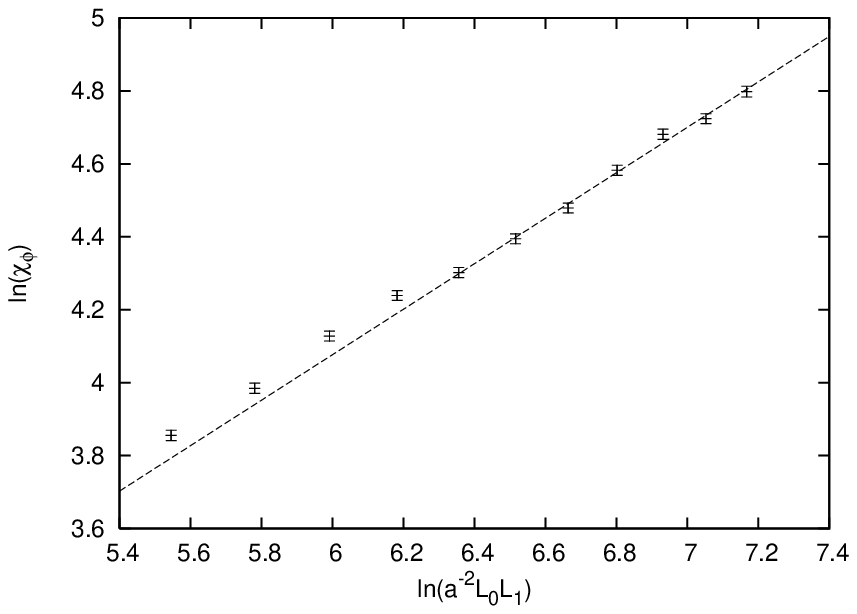}
\caption{$\ln(\chi_\phi)$ as a function of $\ln(a^{-2}L_0L_1)$. UV
subtraction~\eqref{eq:(6.4)} with $n_0\times n_1=3\times3$ is made. The broken
line is a linear fit with $1-h-\Bar{h}=0.624$.}
\label{fig:6}
\end{figure}

In~Eq.~\eqref{eq:(6.2)}, the integral over~$x$ is performed for all~$x$
including the coincidence point~$x=0$. This might be physically unnatural
because the coincidence point~$x=0$ would suffer from ambiguity associated with
the UV regularization. On the other hand, true long-distance physics should
be independent of such a UV ambiguity. To avoid such UV ambiguity and to
estimate how much our determination depends on the UV prescription, we could
define the susceptibility with the UV part subtracted~\cite{Kawai:2010yj}:
\begin{align}
   \chi_\phi&\equiv
   \frac{1}{a^2}\int_{L_0L_1-n_0n_1a^2}d^2x\left\langle A(x)A^*(0)\right\rangle
\notag\\
   &=\frac{1}{a^2L_0L_1}\left\langle\left|\Tilde{A}(0)\right|^2\right\rangle
\notag\\
   &\qquad{}
   -\frac{1}{a^2(L_0L_1)^2}\sum_p
   \frac{2}{p_0}\sin\left(\frac{p_0an_0}{2}\right)
   \frac{2}{p_1}\sin\left(\frac{p_1an_1}{2}\right)
   \left\langle\left|\Tilde{A}(p)\right|^2\right\rangle,
\label{eq:(6.4)}
\end{align}
where the integral has been defined by extracting a region $a^2n_0\times n_1$
containing the coincidence point~$x=0$; in this expression it is understood
that $(2/p_\mu)\sin(p_\mu an_\mu/2)=an_\mu$ for~$p_\mu=0$. Figure~\ref{fig:6} is
the result with the UV subtraction with~$n_0\times n_1=3\times3$ (this is a
choice identical to~Ref.~\cite{Kawai:2010yj}).
Our linear fits are summarized in~Table~\ref{table:3}. By adopting the average
of the second and the third rows in~Table~\ref{table:3} as the central value
and estimating a systematic error associated with the fitting range and the UV
ambiguity by variations in~Table~\ref{table:3}, we quote
\begin{equation}
   1-h-\Bar{h}=0.616(25)(13).
\label{eq:(6.5)}
\end{equation}
This value is somewhat smaller than the value obtained
in~Ref.~\cite{Kawai:2010yj}, but still consistent with the expected exact value
$1-h-\Bar{h}=2/3=0.666\dots$ within~$1.3\sigma$.

\section{Central charge from a supercurrent correlator}
The central charge~$c$, defined by the Virasoro anomaly
\begin{equation}
   \left\langle T_{zz}(x)T_{zz}(0)\right\rangle
   =\frac{1}{8\pi^2}\frac{c/2}{z^4},
\label{eq:(7.1)}
\end{equation}
where $T_{zz}$ is the holomorphic part of the energy-momentum tensor, is a
fundamental quantity that characterizes a conformal field theory (CFT). It is
thus of great interest whether we can measure this quantity~$c$ in the IR
region from the present numerical simulation and obtain further support for the
conjectured correspondence between the WZ model and SCFT. A salient feature of
the present formulation is its nice symmetry properties, including
translational invariance. Thus, we may define the energy-momentum tensor as a
Noether current associated with the translational invariance in a regularized
framework.

In an $\mathcal{N}=2$ SCFT, the central charge~$c$ appears also in the
correlation function between the holomorphic $U(1)$ currents: 
\begin{equation}
   \left\langle J_z(x)J_z(0)\right\rangle
   =\frac{1}{8\pi^2}\frac{c/3}{z^2}.
\label{eq:(7.2)}
\end{equation}
In the LG description, the $U(1)$ symmetry in the $\mathcal{N}=2$ SCFT is
identified with the $U(1)_R$ symmetry of the WZ model with a
(quasi-)homogeneous superpotential (see below) and this $U(1)_R$ is also
exactly preserved in the present formulation. We can thus define a conserved
$U(1)_R$ current whose appropriate component is identified with~$J_z$.

In preliminary numerical study of the above two-point functions of
\emph{bosonic} operators ($T_{zz}$ for~Eq.~\eqref{eq:(7.1)} and $J_z$
for~Eq.~\eqref{eq:(7.2)}), however, we found that signals are generally very
noisy and reliable fits for the central charge would be rather
difficult.\footnote{We stress that this preliminary study was quite incomplete.
In particular, we have not seriously considered the necessity of subtracting
some disconnected parts from correlation functions. Clearly additional study is
needed to conclude something about the use of those correlation functions in
computation of the central charge.} So, in what follows, we adopt a different
approach that employs a two-point function of the $\mathcal{N}=2$
superconformal currents:
\begin{equation}
   \left\langle G_z^+(x)G_z^-(0)\right\rangle
   =\frac{1}{8\pi^2}\frac{2c/3}{z^3}.
\label{eq:(7.3)}
\end{equation}
Since the operators $G_z^\pm$ are \emph{fermionic}, there is no
``disconnected diagram'' which could contribute to Eq.~\eqref{eq:(7.3)} and we
expect a clear signal from numerical simulation. In fact the following results
seem to be in accord with this naive expectation.

Now, since our present formulation possesses exact SUSY, we can define a
conserved supercurrent. To obtain this, we consider the localized SUSY
transformation (for a generic field~$\varphi$)
\begin{equation}
   \delta\Tilde{\varphi}(p)
   =\frac{1}{L_0L_1}\sum_q\left[
   \sum_{\alpha=1}^2
   \Tilde{\xi}^\alpha(q)Q_\alpha\Tilde{\varphi}(p-q)
   -\sum_{\Dot{\alpha}=\Dot{1}}^{\Dot{2}}
   \Tilde{\Bar{\xi}}^{\Dot{\alpha}}(q)\Bar{Q}_{\Dot{\alpha}}\Tilde{\varphi}(p-q)
   \right],
\end{equation}
where $\Tilde{\xi}^\alpha(q)$ and~$\Tilde{\Bar{\xi}}^{\Dot{\alpha}}(q)$ are
non-constant Grassmann parameters and $Q_\alpha$ and~$\Bar{Q}_{\Dot{\alpha}}$ are
defined by~Eqs.~\eqref{eq:(5.1)}--\eqref{eq:(5.4)}. Then from the variation of
the action, we define supercurrents $\Tilde{S}_\mu^\pm(p)$
and~$\Tilde{\Bar{S}}_\mu^\pm(p)$ as
\begin{align}
   \delta S&\equiv\frac{1}{L_0L_1}
   \sum_p(-2)\sum_\mu\Bigl[
   \Tilde{\Bar{\xi}}^{\Dot{2}}(-p)(-ip_\mu)\Tilde{S}_\mu^+(p)
   +\Tilde{\xi}^2(-p)(-ip_\mu)\Tilde{S}_\mu^-(p)
\notag\\
   &\qquad\qquad\qquad\qquad\qquad{}
   +\Tilde{\Bar{\xi}}^{\Dot{1}}(-p)(-ip_\mu)\Tilde{\Bar{S}}_\mu^+(p)
   +\Tilde{\xi}^1(-p)(-ip_\mu)\Tilde{\Bar{S}}_\mu^-(p)
   \Bigr].
\label{eq:(7.5)}
\end{align}
By construction, above supercurrents satisfy SUSY WT identities, such as
\begin{equation}
   p_\mu\left\langle\Tilde{S}_\mu^+(p)
   \Tilde{\varphi}_1(q_1)\dots\Tilde{\varphi}_n(q_n)
   \right\rangle
   =\frac{i}{2}\sum_{i=1}^n\left\langle
   \Tilde{\varphi}_1(q_1)\dots\Bar{Q}_{\Dot{2}}\Tilde{\varphi}_i(q_i+p)
   \dots\Tilde{\varphi}_n(q_n)
   \right\rangle.
\end{equation}
This identity would imply that (according to the standard argument) the
current~$\Tilde{S}_\mu^+$ is a correctly normalized operator.

In Eq.~\eqref{eq:(7.5)}, the explicit form of the supercurrents are given by
\begin{align}
   &\Tilde{S}_z^+(p)
   =\frac{1}{L_0L_1}\sum_q\sqrt{2}i(p-q)_z
   \Tilde{A}(p-q)\Tilde{\Bar{\psi}}_{\Dot{2}}(q),
\label{eq:(7.7)}\\
   &\Tilde{S}_{\Bar{z}}^+(p)
   =\frac{1}{L_0L_1}\sum_q
   \frac{1}{\sqrt{2}}W'(\Tilde{A})(p-q)\Tilde{\psi}_1(q),
\label{eq:(7.8)}\\
   &\Tilde{S}_z^-(p)
   =-\frac{1}{L_0L_1}\sum_q\sqrt{2}i(p-q)_z
   \Tilde{A^*}(p-q)\Tilde{\psi}_2(q),
\label{eq:(7.9)}\\
   &\Tilde{S}_{\Bar{z}}^-(p)
   =\frac{1}{L_0L_1}\sum_q
   \frac{1}{\sqrt{2}}W'(\Tilde{A})^*(p-q)\Tilde{\Bar{\psi}}_{\Dot{1}}(q),
\label{eq:(7.10)}
\end{align}
and
\begin{align}
   &\Tilde{\Bar{S}}_{\Bar{z}}^+(p)
   =\frac{1}{L_0L_1}\sum_q\sqrt{2}i(p-q)_{\Bar{z}}
   \Tilde{A}(p-q)\Tilde{\Bar{\psi}}_{\Dot{1}}(q),
\label{eq:(7.11)}\\
   &\Tilde{\Bar{S}}_z^+(p)
   =-\frac{1}{L_0L_1}\sum_q
   \frac{1}{\sqrt{2}}W'(\Tilde{A})(p-q)\Tilde{\psi}_2(q),
\label{eq:(7.12)}\\
   &\Tilde{\Bar{S}}_{\Bar{z}}^-(p)
   =-\frac{1}{L_0L_1}\sum_q\sqrt{2}i(p-q)_{\Bar{z}}
   \Tilde{A^*}(p-q)\Tilde{\psi}_1(q),
\label{eq:(7.13)}\\
   &\Tilde{\Bar{S}}_z^-(p)
   =-\frac{1}{L_0L_1}\sum_q
   \frac{1}{\sqrt{2}}W'(\Tilde{A})^*(p-q)\Tilde{\Bar{\psi}}_{\Dot{2}}(q).
\label{eq:(7.14)}
\end{align}
A Noether current is, however, always ambiguous as one can change the
definition as $\Tilde{S}_z^\pm(p)\to\Tilde{S}_z^\pm(p)+p_z\Tilde{X}^\pm(p)$ and
$\Tilde{S}_{\Bar{z}}^\pm(p)\to\Tilde{S}_{\Bar{z}}^\pm(p)%
-p_{\Bar{z}}\Tilde{X}^\pm(p)$ by using certain combinations $\Tilde{X}^\pm(p)$
without affecting the conservation law,
$p_{\Bar{z}}\Tilde{S}_z^\pm(p)+p_z\Tilde{S}_{\Bar{z}}^\pm(p)=0$. To fix this
ambiguity, we required in~Eqs.~\eqref{eq:(7.7)} to~\eqref{eq:(7.14)} that
\begin{equation}
   \Tilde{S}_{\Bar{z}}^+=\Tilde{S}_{\Bar{z}}^-
   =\Tilde{\Bar{S}}_z^+=\Tilde{\Bar{S}}_z^-=0,
\label{eq:(7.15)}
\end{equation}
when $W'=0$. The WZ model with~$W'=0$ is a massless free theory that itself is
an $\mathcal{N}=(2,2)$ SCFT. For this system it is natural to adopt a
supercurrent that obeys the gamma-traceless condition,
\begin{equation}
   \sum_\mu\gamma_\mu
   \begin{pmatrix}
   \Tilde{\Bar{S}}_\mu^-\\\Tilde{S}_\mu^+
   \end{pmatrix}=
   \sum_\mu\gamma_\mu
   \begin{pmatrix}
   \Tilde{\Bar{S}}_\mu^+\\\Tilde{S}_\mu^-
   \end{pmatrix}=0,
\label{eq:(7.16)}
\end{equation}
because this condition is a super-partner of the traceless
condition~$T_{z\Bar{z}}=0$~\cite{Ferrara:1974pz} that is usually assumed in CFT.
Eq.~\eqref{eq:(7.15)} is nothing but Eq.~\eqref{eq:(7.16)} in components.

Thus we had a natural definition of the supercurrents. We then postulate a
correspondence between the components of above supercurrents and the
holomorphic part of the superconformal currents in the IR limit. That is,
\begin{equation}
   \Tilde{S}_z^+(p)\to\Tilde{G}_z^+(p),\qquad
   \Tilde{S}_z^-(p)\to\Tilde{G}_z^-(p).
\label{eq:(7.17)}
\end{equation}
Our reasoning for this correspondence is as follows:
\begin{itemize}
\item[(i)] The conservation law of the supercurrents in the coordinate space
yields $\partial_{\Bar{z}}S_z^\pm(x)+\partial_zS_{\Bar{z}}^\pm(x)=0$. In the IR
limit, the derivative of the superpotential~$W'(A)$ is expected to become
irrelevant (see, for example, Section~14.4 of~Ref.~\cite{Hori:2003ic}) so
$S_{\Bar{z}}^\pm(x)$, from explicit forms~\eqref{eq:(7.8)}
and~\eqref{eq:(7.10)}, could be neglected in the IR limit. Then the
conservation law implies that $S_z^\pm(x)$ are holomorphic functions, a correct
property of~$G_z^\pm(x)$.
\item[(ii)] The WZ model with a homogeneous superpotential $W=\lambda\Phi^n/n$
possesses the $U(1)_R$~symmetry
\begin{align}
   &A\to e^{(2/n)i\theta}A,\qquad
   \psi_\alpha\to e^{-(1-2/n)i\theta}\psi_\alpha,
\notag\\
   &\Bar{\psi}_{\Dot{\alpha}}\to e^{(1-2/n)i\theta}\Bar\psi_{\Dot{\alpha}},\qquad
   F\to e^{-(2-2/n)i\theta}F,
\label{eq:(7.18)}
\end{align}
which is, in the IR, identified with a $U(1)$ symmetry generated by a
\emph{sum} of two $U(1)$ charges $(q,\Bar{q})$ in the left- and right-moving
$\mathcal{N}=2$ superconformal algebras (see, for example, Section~19.4
of~Ref.~\cite{Polchinski:1998rr}). Thus, in~Eq.~\eqref{eq:(7.18)}, the $U(1)_R$
charge of the scalar field~$A$ is assigned~$2/n$, because $A$ is identified
with a chiral primary field in the $A_{n-1}$~model with the $U(1)$
charges~$(q,\Bar{q})=(1/n,1/n)$. Then, assuming that $\Tilde{S}_z^+$
in~Eq.~\eqref{eq:(7.7)} possesses $U(1)$ charges $(q,0)$ in the IR, we see from
the $U(1)$ charge assignment in~Eq.~\eqref{eq:(7.18)} that $q=+1$, a correct
$U(1)$ charge of~$G_z^+$. Similarly, for $\Tilde{S}_z^-$
in~Eq.~\eqref{eq:(7.9)}, we have~$q=-1$, a correct $U(1)$ charge of~$G_z^-$.
\item[(iii)] The overall normalization of the supercurrents has been fixed such
that Eq.~\eqref{eq:(7.3)} is reproduced under correspondence~\eqref{eq:(7.17)}
for the massless free theory ($W'=0$). This theory itself is an
$\mathcal{N}=(2,2)$ SCFT whose left-moving sector possesses the central
charge~$c=3$ (one complex scalar and two Majorana-Weyl fermions).
\end{itemize}
Thus, under identification~\eqref{eq:(7.17)}, our procedure is as follows: we
numerically compute the two-point function of the supercurrents in the momentum
space
\begin{equation}
   \left\langle\Tilde{S}_z^+(p)\Tilde{S}_z^-(-p)\right\rangle.
\label{eq:(7.19)}
\end{equation}
Then, in view of correspondence~\eqref{eq:(7.17)}, we compare this function in
the IR region\footnote{Note that the coupling constant~$\lambda$
in~Eq.~\eqref{eq:(1.1)} is a unique physical dimensionful parameter in the
present system on~$\mathbb{R}^2$.} $|p|\lesssim\lambda$ with the two-point
function of the superconformal currents in the momentum space
\begin{align}
   &L_0L_1\int d^2x\,e^{-ipx}
   \left\langle G_z^+(x)G_z^-(0)\right\rangle
\notag\\
   &=L_0L_1\frac{-ic}{48\pi}\frac{\partial^3}{\partial p_{\Bar{z}}^3}
   \frac{p^2}{\delta^2}K_2(|p|\delta)
   \xrightarrow{|p|\delta\to0}
   L_0L_1\frac{ic}{24\pi}\frac{p_z^2}{p_{\Bar{z}}},
\label{eq:(7.20)}
\end{align}
where we have used Eq.~\eqref{eq:(7.3)} that is, strictly speaking, an
expression valid only on~$\mathbb{R}^2$. In deriving the last expression, we
regularized the singularity in the integrand at~$x=0$ by setting
$1/z^3=(\Bar{z})^3/(x^2)^3\to(\Bar{z})^3/(x^2+\delta^2)^3$. Note that the last
low-momentum behavior (i.e., the IR physics) in~Eq.~\eqref{eq:(7.20)} is
independent of the regularization parameter~$\delta$.

\begin{figure}[htb]
\centering
\includegraphics[width=120mm]{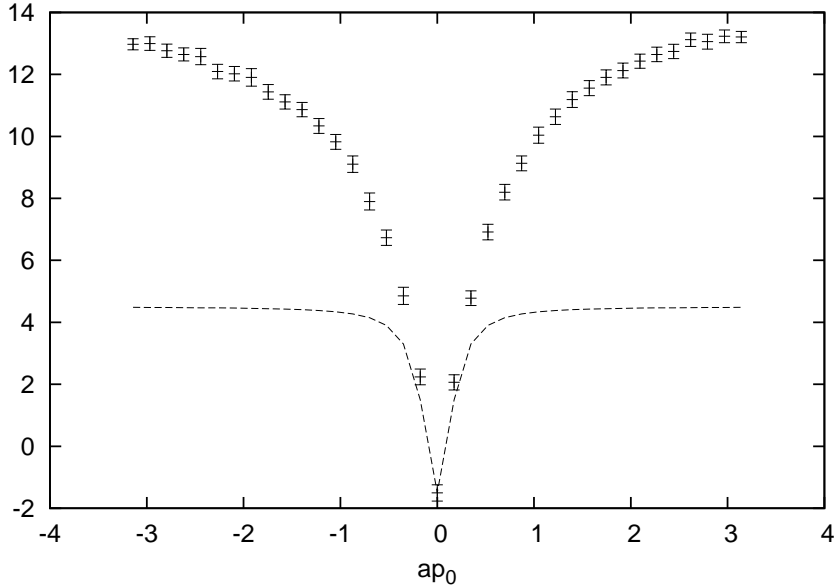}
\caption{The real part of the correlation function~\eqref{eq:(7.19)} as a
function of~$ap_0$ along the line $ap_1=\pi/18\sim0.1745$;
$N_0\times N_1=36\times36$. The broken line is the real part of
function~\eqref{eq:(7.20)} with~$c=1$.}
\label{fig:7}
\end{figure}

\begin{figure}[htb]
\centering
\includegraphics[width=120mm]{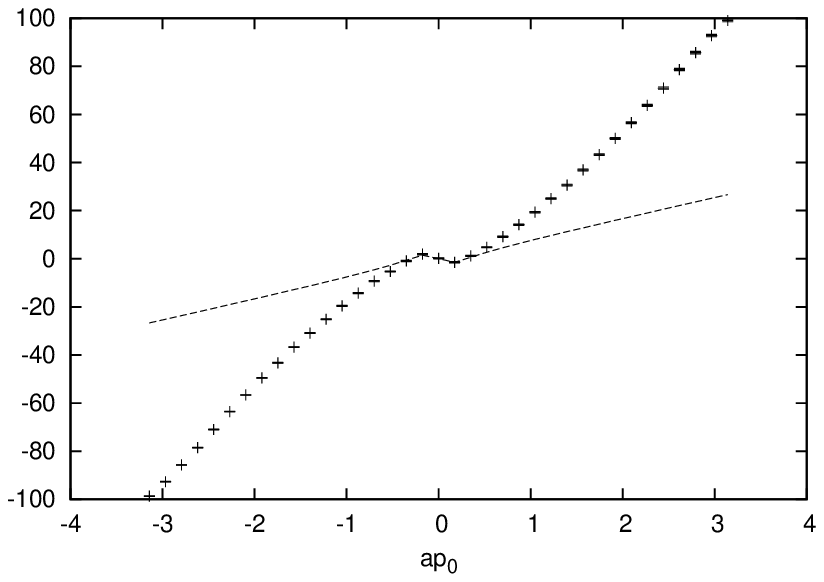}
\caption{The imaginary part of the correlation function~\eqref{eq:(7.19)} as a
function of~$ap_0$ along the line $ap_1=\pi/18\sim0.1745$;
$N_0\times N_1=36\times36$. The broken line is the imaginary part of
function~\eqref{eq:(7.20)} with~$c=1$.}
\label{fig:8}
\end{figure}

For the above computation, we used only data with the finest
grid~$N_0\times N_1=36\times36$. In~Fig.~\ref{fig:7}, we depicted the real
part of correlation function~\eqref{eq:(7.19)} as a function of~$ap_0$ along
the line~$ap_1=\pi/18\sim0.1745$.
The broken line is the real part of function~\eqref{eq:(7.20)} with~$c=1$; it
agrees well with the data for $|ap_0|\leq\pi/18\sim0.1745<0.3$, a good
indication for~$c\sim1$ in the IR region. Fig.~\ref{fig:8} is the same
as~Fig.~\ref{fig:7}, but for the imaginary part.
The broken line is the imaginary part of function~\eqref{eq:(7.20)} with~$c=1$;
this time it agrees well with the data for $|ap_0|\leq2\pi/18\sim0.349
\simeq0.3$.

\begin{figure}[htb]
\centering
\includegraphics[width=120mm]{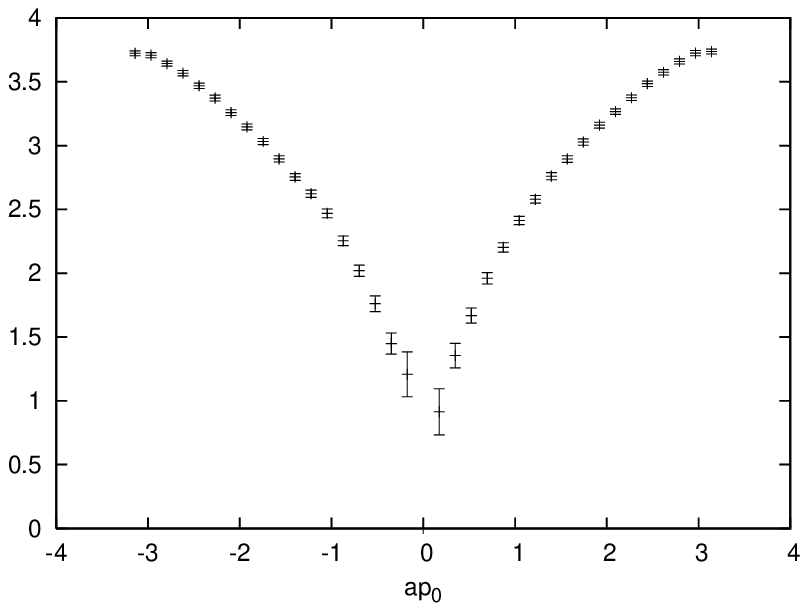}
\caption{The real part of the ratio of the correlation
function~\eqref{eq:(7.19)} and the function~$L_0L_1(i/24\pi)p_z^2/p_{\Bar{z}}$ as
a function of~$ap_0$ along the line~$ap_1=0$; $N_0\times N_1=36\times36$. The
origin~$p=0$ is excluded from the plot because $p_z^2/p_{\Bar{z}}$ is singular
at the origin.}
\label{fig:9}
\end{figure}

\begin{figure}[htb]
\centering
\includegraphics[width=120mm]{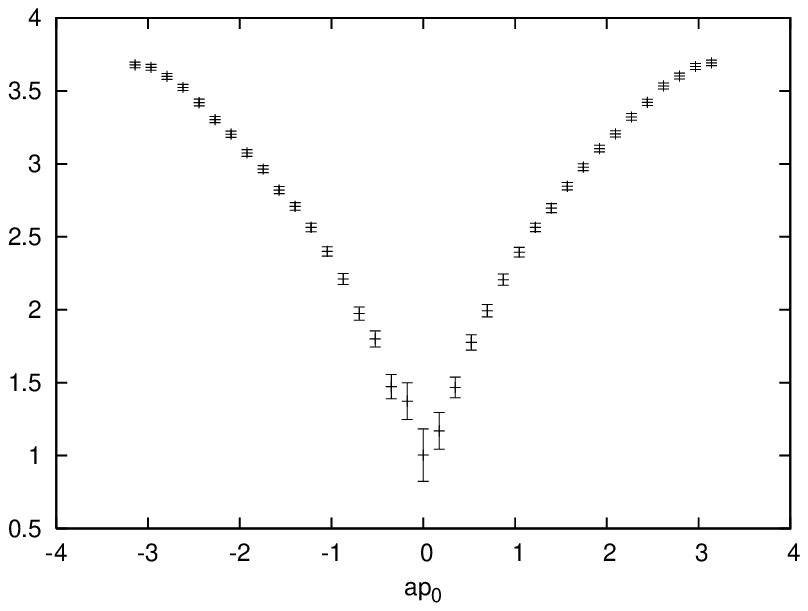}
\caption{The real part of the ratio of the correlation
function~\eqref{eq:(7.19)} and the function~$L_0L_1(i/24\pi)p_z^2/p_{\Bar{z}}$ as
a function of~$ap_0$ along the line~$ap_1=\pi/18\sim0.1745$;
$N_0\times N_1=36\times36$.}
\label{fig:10}
\end{figure}

\begin{figure}[htb]
\centering
\includegraphics[width=120mm]{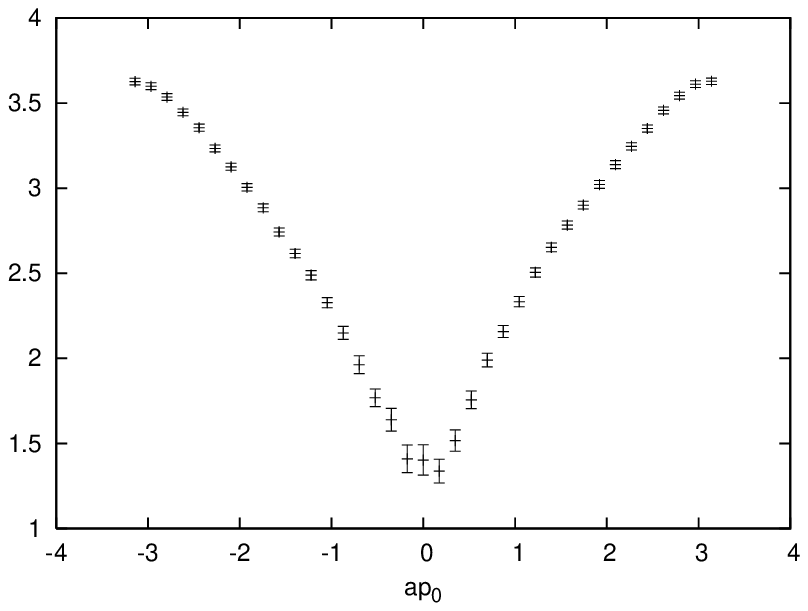}
\caption{The real part of the ratio of the correlation
function~\eqref{eq:(7.19)} and the function~$L_0L_1(i/24\pi)p_z^2/p_{\Bar{z}}$ as
a function of~$ap_0$ along the
line~$ap_1=2\pi/18\sim0.349$; $N_0\times N_1=36\times36$.}
\label{fig:11}
\end{figure}

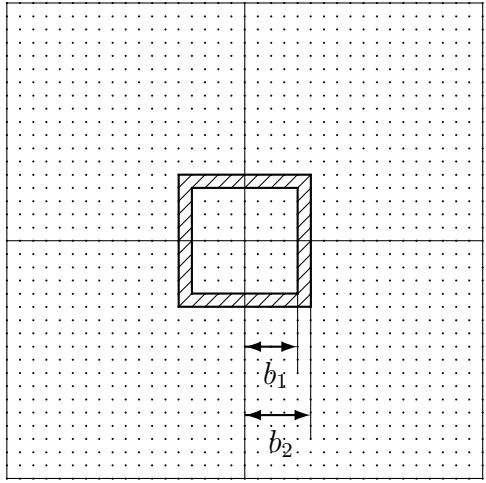
\begin{figure}[htb]
\centering
\begin{picture}(180,180)(0,-10)
\put(90,90){\line(1,0){90}}
\put(90,90){\line(-1,0){90}}
\put(90,90){\line(0,1){90}}
\put(90,90){\line(0,-1){90}}
\multiput(65,65)(5,0){10}{\line(1,1){5}}
\multiput(65,70)(0,5){9}{\line(1,1){5}}
\multiput(70,110)(5,0){9}{\line(1,1){5}}
\multiput(110,70)(0,5){8}{\line(1,1){5}}
\put(110,70){\line(0,-1){30}}
\put(115,75){\line(0,-1){60}}
\put(97,36){$b_1$}
\put(99,10){$b_2$}
\put(90,90){\oval[0](180,180)}
\multiput(0,0)(0,5){37}{%
\multiput(0,0)(5,0){37}{\circle*{0.8}}%
}
\thicklines
\put(90,90){\oval[0](40,40)}
\put(90,90){\oval[0](50,50)}
\put(90,50){\vector(1,0){20}}
\put(90,50){\vector(-1,0){0}}
\put(90,24){\vector(1,0){25}}
\put(90,24){\vector(-1,0){0}}
\end{picture}
\caption{The fitting region $R(b_1,b_2)$ on the $N_0\times N_1=36\times36$
momentum grid. Data points at the boundaries as well as data points inside
the shaded region are used for a constant fit.}
\label{fig:12}
\end{figure}

Our fit for~$c$ proceeds as follows: We first consider the real part of the
ratio of correlation function~\eqref{eq:(7.19)} to the
function~$L_0L_1(i/24\pi)p_z^2/p_{\Bar{z}}$ appearing in~Eq.~\eqref{eq:(7.20)}.
For illustration, the real part of this ratio along several constant $ap_1$
lines are depicted in~Figs.~\ref{fig:9}, \ref{fig:10} and~\ref{fig:11}.
We see that actually the ratio is close to~$1$ around the origin in the
momentum space $|p|\sim0$. Then, to use data points with a fixed energy scale,
we define a fitting region~$R(b_1,b_2)$ on the momentum grid as illustrated
in~Fig.~\ref{fig:12}.
Then we carry out a $\chi^2$~fit by a constant for data points in the fitting
region~$R(b_1,b_2)$. Since the IR region is characterized by the condition
$|ap|\lesssim a\lambda=0.3$ and the lowest non-zero momentum in the present
$N_0\times N_1=36\times36$ momentum grid is $ap_\mu=\pi/18\sim0.1745$, we
regard the fitting region $R(1,1)$, or at most $R(1,2)$, as the IR region;
see~Table~\ref{table:4}.
\begin{table}
\caption{Results of a $\chi^2$ fit used for a determination of the central
charge~$c$ in~\eqref{eq:(7.21)}.}
\label{table:4}
\begin{center}
\begin{tabular}{lrll}
\hline
Fitting region & Number of data points & $\chi^2/\text{d.o.f.}$ & $c$\\
\hline
$R(1,1)$ & $8$ & $1.93$ & $1.09(14)$\\
$R(1,2)$ & $24$ & $3.50$ & $1.40(8)$\\
\hline
\end{tabular}
\end{center}
\end{table}
From the numbers in the table, we estimate
\begin{equation}
   c=1.09(14)(31),
\label{eq:(7.21)}
\end{equation}
as the central charge in the IR region. Our estimate reproduces the conjectured
value $c=1$ very well.

\begin{figure}[htb]
\centering
\includegraphics[width=120mm]{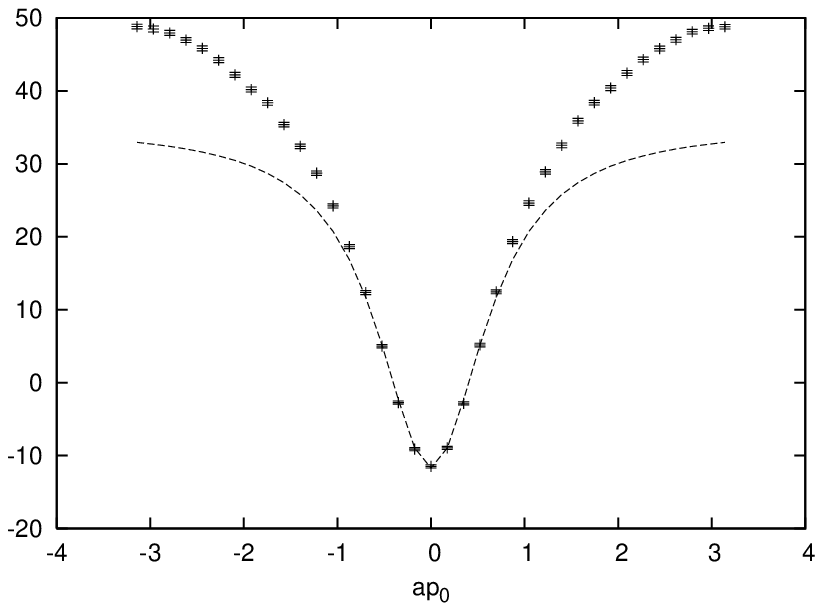}
\caption{The real part of the correlation function~\eqref{eq:(7.19)} as a
function of~$ap_0$ along the line $ap_1=4\pi/18\sim0.698$;
$N_0\times N_1=36\times36$. The broken line is the real part of
function~\eqref{eq:(7.20)} with~$c=1.95$, that was obtained by a constant fit
in the fitting region~$R(4,4)$ ($\chi^2/\text{d.o.f.}=1.58$).}
\label{fig:13}
\end{figure}

\begin{figure}[htb]
\centering
\includegraphics[width=120mm]{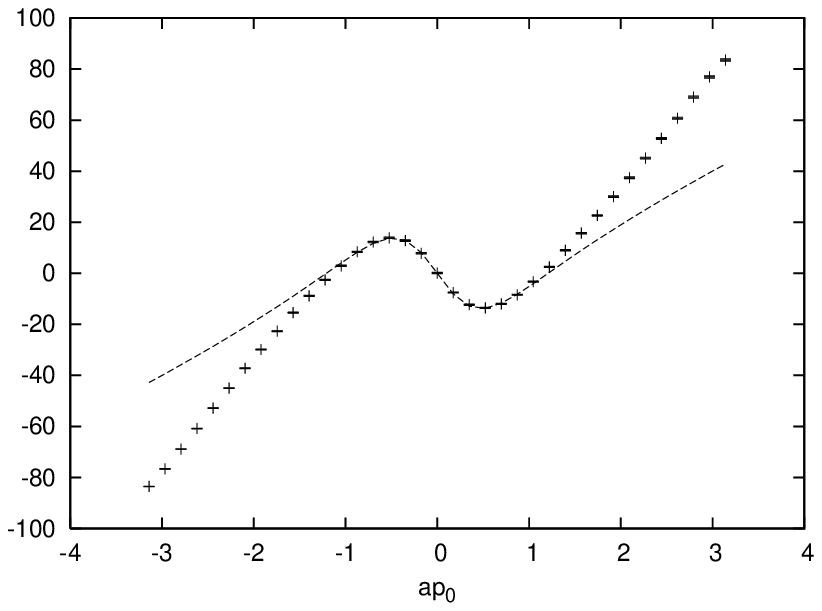}
\caption{The imaginary part of the correlation function~\eqref{eq:(7.19)} as a
function of~$ap_0$ along the line $ap_1=4\pi/18\sim0.698$;
$N_0\times N_1=36\times36$. The broken line is the imaginary part of
function~\eqref{eq:(7.20)} with~$c=1.95$.}
\label{fig:14}
\end{figure}

\begin{figure}[htb]
\centering
\includegraphics[width=120mm]{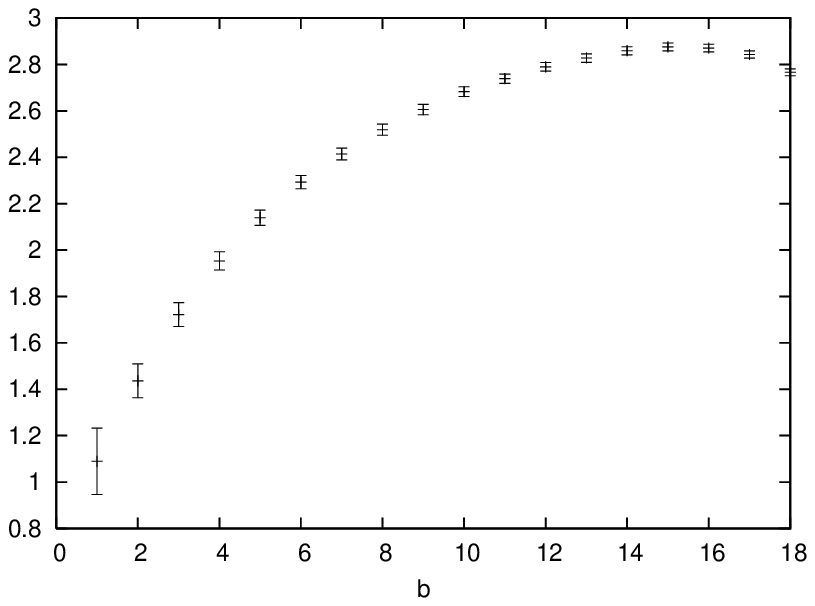}
\caption{``Effective $c$'' obtained from a constant fit in the fitting region
$R(b,b)$.}
\label{fig:15}
\end{figure}

Although strictly speaking expression~\eqref{eq:(7.20)} is meaningful only
for very small momenta and the LG description is expected to be valid only for
the low-energy region $|ap|\lesssim a\lambda=0.3$, we found that it is
nevertheless interesting to repeat the above procedure for ``intermediate''
energy $|ap|\gtrsim a\lambda=0.3$ or even ``high'' energy $|ap|\sim\pi$
regions. Figures~\ref{fig:13} and~\ref{fig:14} is a result of the fitting in an
``intermediate'' energy region $|ap|\sim0.7$.
We observe a rather good fit with an ``effective $c$'', $c\sim2$. We may repeat
such a fit by changing the parameter~$b$ in~$R(b,b)$ which roughly corresponds
to the energy scale. The result of such fits as a function of~$b$ are shown is
Fig.~\ref{fig:15}.
It is interesting that, going from the UV side to the IR side, the plot goes
from~$c\sim3$, the central charge corresponding to the massless free theory,
and decreases to~$c\sim1$, the central charge of the $A_2$~model. Although one
should not take this plot so seriously because $\chi^2$ becomes quite large for
large~$b$ (for example, $\chi^2/\text{d.o.f.}\sim19$ for~$b=8$), it is still
interesting to note an analogue of the Zamolodchikov
$C$~function~\cite{Zamolodchikov:1986gt} in this plot; the $C$~function
monotonically decreases along the renormalization group flow from UV to IR and
coincides with the central charge at a fixed point. It is an interesting
problem to clarify the underlying reason for this similarity
of~Fig.~\ref{fig:15} with the $C$~function.

\section{Conclusion}
In this paper, we carried out a non-perturbative numerical study of low-energy
physics of the 2D WZ model with the massless cubic superpotential. We obtained
the critical exponent $1-h-\Bar{h}$ and the central charge $c$ which are
consistent with the conjectured emergence of a non-trivial $\mathcal{N}=(2,2)$
SCFT, thus provided further support for the LG description. Our results
indicate that our supersymmetric non-perturbative formulation of the WZ model
is working, although there has been an issue concerning the locality in this
formulation. For further improvement of these results, better observables
which yield less systematic errors should be investigated. Also, we want to
generalize the present analysis to higher critical LG models.

We would like to thank Michael G.~Endres, Kazuo Fujikawa, Masafumi Fukuma,
Daisuke Kadoh, Hikaru Kawai, Akitsugu Miwa, and Tsuneo Uematsu for discussions
and comments. We are quite indebted to Hiroki Kawai and Yoshio Kikukawa for
discussions and detailed explanation on the simulation
in~Ref.~\cite{Kawai:2010yj} and to Michael G.~Endres for a careful reading of
the manuscript. Our numerical calculations were carried out by using the RIKEN
Integrated Cluster of Clusters (RICC) facility. The work of H.S.\ is supported
in part by a Grant-in-Aid for Scientific Research, 22340069 and~23540330.

%% The Appendices part is started with the command \appendix;
%% appendix sections are then done as normal sections
%% \appendix

%% \section{}
%% \label{}

%% References
%%
%% Following citation commands can be used in the body text:
%% Usage of \cite is as follows:
%%   \cite{key}         ==>>  [#]
%%   \cite[chap. 2]{key} ==>> [#, chap. 2]
%%

%% References with bibTeX database:

\bibliographystyle{elsarticle-num}
\bibliography{<your-bib-database>}

%% Authors are advised to submit their bibtex database files. They are
%% requested to list a bibtex style file in the manuscript if they do
%% not want to use elsarticle-num.bst.

%% References without bibTeX database:

\end{document}